\documentclass[lettersize,journal]{IEEEtran}
\usepackage{amsmath,amsfonts}
\usepackage{algorithmic}
\usepackage{algorithm}
\usepackage{array}
\usepackage[caption=false,font=normalsize,labelfont=sf,textfont=sf]{subfig}
\usepackage{textcomp}
\usepackage{stfloats}
\usepackage{url}
\usepackage{verbatim}
\usepackage{graphicx}
\usepackage{cite}
\usepackage{makecell}
\usepackage{amssymb}
\usepackage{threeparttable}
\usepackage[colorlinks=true, linkcolor=blue, citecolor=blue]{hyperref}
\usepackage{tabularx}  

\usepackage{tikz,xcolor,hyperref} 
\definecolor{lime}{HTML}{A6CE39}
\DeclareRobustCommand{\orcidicon}{
\begin{tikzpicture}
\draw[lime, fill=lime] (0,0) 
circle [radius=0.16] 
node[white] {{\fontfamily{qag}\selectfont \tiny ID}};    \draw[white, fill=white] (-0.0625,0.095) 
circle [radius=0.007];    \end{tikzpicture}
\hspace{-2mm}}
\foreach \x in {A, ..., Z}{
\expandafter\xdef\csname orcid\x\endcsname{\noexpand\href{https://orcid.org/\csname orcidauthor\x\endcsname}{\noexpand\orcidicon}}}

\begin{document}
\setcounter{page}{504}

\title{A Hardware-Friendly Shuffling Countermeasure Against Side-Channel Attacks for Kyber}

\author{Dejun Xu\orcidA{}, Kai Wang\orcidB{}, and Jing Tian\orcidC{}, \textit{Member, IEEE}

\thanks{
This work was supported in part by the Natural Science Foundation of Jiangsu Province of China under Grant BK20243038, in part by the Key Research Plan of Jiangsu Province of China under Grant BE2022098, and in part by the Young Elite Scientists Sponsorship Program by the China Association for Science and Technology under Grant 2023QNRC001. \textit{(Corresponding author: Jing Tian.)}

The authors are with the School of Integrated Circuits, Nanjing University, Suzhou, 215163, China (e-mail: tianjing@nju.edu.cn).}}

\markboth{IEEE TRANSACTIONS ON CIRCUITS AND SYSTEMS—II: EXPRESS BRIEFS, VOL. 72, NO. 3, MARCH 2025}{XU et al.: HARDWARE-FRIENDLY SHUFFLING COUNTERMEASURE AGAINST SCAs FOR KYBER} 

\maketitle

\begin{abstract}
CRYSTALS-Kyber has been standardized as the only key-encapsulation mechanism (KEM) scheme by NIST to withstand attacks by large-scale quantum computers. However, the side-channel attacks (SCAs) on its implementation are still needed to be well considered for the upcoming migration. In this brief, we propose a secure and efficient hardware implementation for Kyber by incorporating a novel compact shuffling architecture. First of all, we modify the Fisher-Yates shuffle to make it more hardware-friendly. We then design an optimized shuffling architecture for the well-known open-source Kyber hardware implementation to enhance the security of all known and potential side-channel leakage points. Finally, we implement the modified Kyber design on FPGA and evaluate its security and performance. The security is verified by conducting correlation power analysis (CPA) and test vector leakage assessment (TVLA) on the hardware. Meanwhile, FPGA place-and-route results show that the proposed design reports only 8.7\% degradation on the hardware efficiency compared with the original unprotected version, much better than existing hardware hiding schemes.
\end{abstract}


\section{Introduction}
\IEEEPARstart{W}{ith} the rapid development of quantum computing, modern cryptography faces severe threats.  Since 2016, NIST has held a post-quantum cryptography (PQC) standardization competition, receiving global proposals. After three rounds, CRYSTALS-Kyber (Kyber) was chosen as the only key-encapsulation mechanism (KEM) protocol~\cite{alagic2022status}. Recently, NIST issued FIPS 203, a standard for module-lattice-based key-encapsulation mechanism (ML-KEM) based on CRYSTALS-Kyber. As of now, research on the efficient hardware implementations of ML-KEM/Kyber has attained a relatively mature stage~\cite{zhang2024super,kim2024configurable}. Nevertheless, during the impending shift from existing conventional cryptographic algorithms to ML-KEM/Kyber, it is equally crucial to assess their security against physical attacks, particularly side-channel attacks (SCAs) and fault injection attacks (FIAs).

SCAs recover keys by acquiring power consumption or electromagnetic emanations from cryptographic devices, and typically employ countermeasures such as hiding or masking to defend against them. FIAs, on the other hand, aim to recover keys by introducing faults into cryptographic devices during their operation. Error detection mechanisms can effectively identify and respond to the injection of these faults. On the SCAs side, Tosun \textit{et al}.~\cite{tosun2024zero} successfully recovered the full key of Kyber from the polynomial multiplication on Kyber's software implementations. Zhao \textit{et al}.~\cite{zhao2023side} recently pointed out that there are three side-channel leakage points in Kyber's decryption procedure, and successfully completed SCAs on Kyber's hardware implementations. On the FIAs side, Ni \textit{et al}.~\cite{ni2024bitstream} successfully implemented bitstream fault injection attacks on Kyber's hardware implementations. In addition, combined fault and power attacks also pose a significant threat to lattice-based cryptographic algorithms.

However, defense work for Kyber, especially on hardware platform, is still far from sufficient. Sarker \textit{et al}.~\cite{sarker2022error} proposed the error detection architectures for hardware/software co-design approaches of number theoretic transform (NTT). Zhao \textit{et al}.~\cite{zhao2023side} and Kamucheka \textit{et al}.~\cite{kamucheka2022masked} proposed two different hardware masking schemes, both of which introduce significant area overhead. On the other hand, two hiding schemes on hardware platform are provided in~\cite{moraitis2023securing} and~\cite{jati2024configurable}. In~\cite{moraitis2023securing}, Morait \textit{et al}. used the method of randomized clock and duplication to reduce the energy correlation during the operation of Kyber. In~\cite{jati2024configurable}, Jati \textit{et al}. adopted the methodology of adding random delays and address \& instruction shuffling to reduce the energy correlation. Nevertheless, these two hiding schemes result in high resource overhead and clock cycle overhead, respectively. In this brief, we try to propose a hiding protection method against SCAs for Kyber's hardware implementation by taking the performance and security both into consideration.

This work provides a good tradeoff between security enhancement and hardware efficiency. Specifically, we modify the Fisher-Yates shuffle to make it more hardware-friendly. Based on the algorithm, we propose an optimized shuffling architecture and apply it to the open-source hardware implementation of Kyber from~\cite{xing2021compact}. We shuffle the sub-operation orders of all the potential side-channel leakage points during Kyber's decryption procedure. The proposed design shows only an 8.7\% efficiency drop and surpasses current hiding schemes.

\section{Preliminaries}
\subsection{Related Work on Shuffling}
As an efficient and effective countermeasure, shuffling is widely used in various cryptographic implementations against SCAs. In \cite{ravi2020configurable}, Ravi \textit{et al}. proposed three variants of the shuffling countermeasure with varying granularity for the NTT on software platform. In \cite{jati2024configurable,chen2022low,zijlstra2019fpga}, three different shuffling countermeasures on the hardware platform are proposed, as shown in Table \ref{tab:shuffling}. The schemes from \cite{jati2024configurable,chen2022low} require lots of storage space to achieve high security as their random permutations are generated outside of the hardware and preloaded into storage in advance. Therefore, they require lots of storage space to achieve high security. Differently,  Zijlstra \textit{et al}.~\cite{zijlstra2019fpga} used the random permutation generator (RPG) from~\cite{bayrak2012architecture} to protect the lattice-based cryptographic schemes. However, the RPG they proposed is quite costly, making it difficult to meet the low-cost requirements.

\begin{table}[!t]
\caption{Comparison with Existing Hardware Shuffling Methods\label{tab:shuffling}}
\centering
\renewcommand{\arraystretch}{1.4}   
\setlength{\tabcolsep}{4.5pt}    
\begin{tabular}{|c|c|c|c|}
\hline
Work & \makecell{Huge Space \\ of Permutations} & \makecell{No Dynamic \\ Input Required} & \makecell{No Pre-Storage \\ Required}\\
\hline
TECS~\cite{jati2024configurable} & $\times$ & $\times$ & $\times$\\
\hline
TCAD~\cite{chen2022low} & $\times$ & $\times$ & $\times$\\
\hline
INDOCRYPT~\cite{zijlstra2019fpga} & $\checkmark$ & $\times$ & $\checkmark$\\
\hline
This work & $\checkmark$ & $\checkmark$ & $\checkmark$\\
\hline
\end{tabular}
\end{table}

\subsection{Kyber and Side-Channel Leakage Points in Kyber}
Kyber is a CCA-secure protocol whose security is based on the module learning with errors (MLWE) problem. The core operations of Kyber are polynomial multiplications over the polynomial ring $\mathcal{R}_q=\mathbb{Z}_q/(X^n+1)$, where $n=256$, $q=3329$, and $k = 2$, $3$, and $4$. The whole Kyber protocol consists of three procedures, which are key generation, encryption, and decryption. The side-channel leakage points of the secret key are mainly concentrated in Kyber's decrypting procedure, and the main operations of this procedure are as follows:
\begin{equation}
\label{1}
m\leftarrow Compress_q(v-INTT(\boldsymbol{\hat{s}}^T\circ NTT(\boldsymbol{u})),1).
\end{equation}
They include NTT, point-wise multiplication (PWM), inverse NTT (INTT), subtraction, and compress, where $m$ denotes the decrypted message, $\boldsymbol{\hat{s}}$ denotes the unpacked secret key, $\boldsymbol{u}$ and $v$ denote the unpacked ciphertext. NTT, PWM, and INTT are used to reduce the complexity of polynomial multiplications from $\mathcal{O}(n^2)$ to $\mathcal{O}(n\cdot logn)$.

According to~\cite{zhao2023side}, there are three side-channel leakage points in Kyber's decryption procedure (hardware) as shown in~\eqref{2}. 
\begin{equation}
\label{2}
\left\{
\begin{aligned}
{\rm point} \ &1: \hat{\boldsymbol{s}}^T\circ \hat{\boldsymbol{u}},\\
{\rm point} \ &2: (\hat{\boldsymbol{s}}^T\circ \hat{\boldsymbol{u}}) \ mod \ q,\\
{\rm point} \ &3: v-(\boldsymbol{s}^T\boldsymbol{u} \ mod \ q).
\end{aligned}
\right.
\end{equation}
The three leakage points lie in PWM, the modular reduction after PWM, and the subtraction after INTT, respectively. In addition, INTT has been successfully attacked in software implementations \cite{hamburg2021chosen}, so we retain the protection of INTT by shuffling the butterfly units using a series of fixed-length random permutations produced by the proposed RPG.

\begin{figure}[!t]
\centering
\includegraphics[width=3.33in]{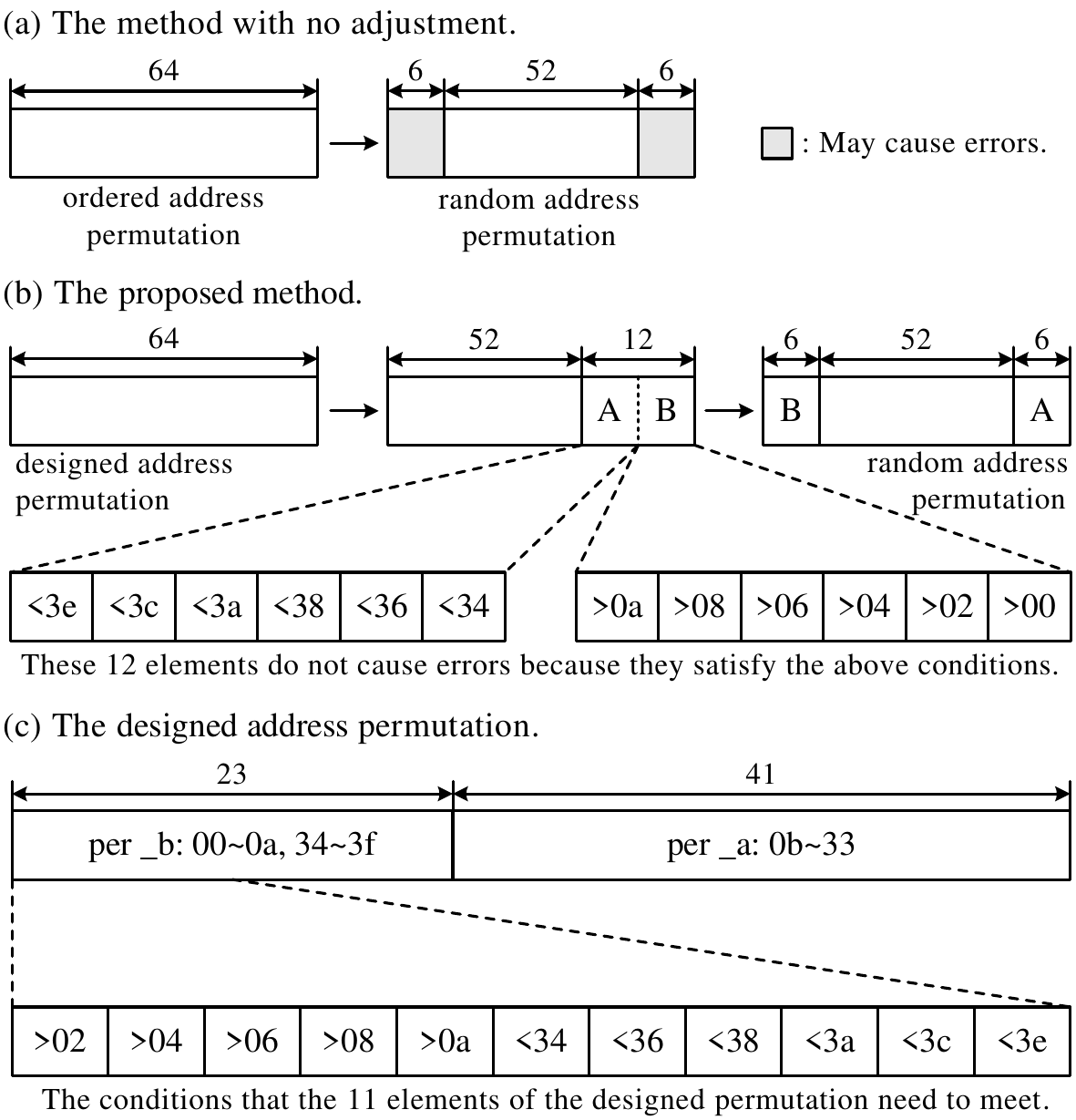}
\caption{The comparison of the shuffling method without and with adjustment.}
\label{method}
\end{figure}

\section{Proposed Shuffling Architecture for Kyber}
Fisher-Yates shuffle is one of the most widely used shuffling algorithms. When it is implemented in hardware, two problems need to be solved. One is to continuously input random seeds and the other is to dynamically reduce the range of indexes. In this work, we have elaborately solved these problems. The key ideas of the improved algorithm will be shown in Algorithm~\ref{alg1}.

\subsection{Random Permutation Generator}
RPG is a module that generates random permutations from an ordered one. There are five different random permutations to be generated, which are $00\sim3f$, $40\sim7f$, $00\sim7f$, $80\sim bf$, and $c0\sim ff$. Since there is a correspondence between these permutations, it is only necessary to generate the permutation  $00\sim3f$, and the other permutations can be obtained by extending it.

Before introducing the hardware architecture, it is necessary to explain that due to the existence of overlapping addresses, it is necessary to impose restrictions on the first six and last six elements of the randomly generated permutations to avoid address conflicts. The key idea is to specially introduce a fixed intermediate permutation and generate the random one in two stages. The comparison of the shuffling methods without and with adjustment are shown in Fig.~\ref{method}. As explained above, if we directly use the random address permutation, there may exist errors in the first six or the last six locations as illustrated in Fig.~\ref{method}(a). To avoid those errors, an intermediate address permutation is specially designed in advance as shown in Fig.~\ref{method}(c). It is divided into two regions, $per\_a$ and $per\_b$, which are defined as: 
\begin{equation}
\label{9}
\left\{
\begin{aligned}
per\_a&=\{33,32,31,...,0d,0c,0b\}, \\
per\_b&=\{34,...,38\}\cup\{00,...,0a\}\cup\{39,...,3f\}.
\end{aligned}
\right.
\end{equation}
Actually, the designed address permutation only needs to restrict those 11 elements and can be in any form. For simplicity, we define it in the above form in this brief. 

In the first stage of shuffling, the source of address permutation is $per\_a$ and twelve addresses are selected based on the proposed random index permutation, which are placed at the rightmost of a new empty permutation. Those selected locations of $per\_a$ are then filled up by the elements of the leftmost of $per\_b$ successively. This stage is to protect from errors. In the second stage of shuffling, the rest random addresses are selected from the remaining 52 elements of $per\_a$ and $per\_b$. After finishing all the selections, the filled permutation is cyclically shifted to the right 6 times and output as the final random address permutation, as shown in Fig.~\ref{method}(b). 

\begin{figure}[!t]
\centering
\includegraphics[width=3.33in]{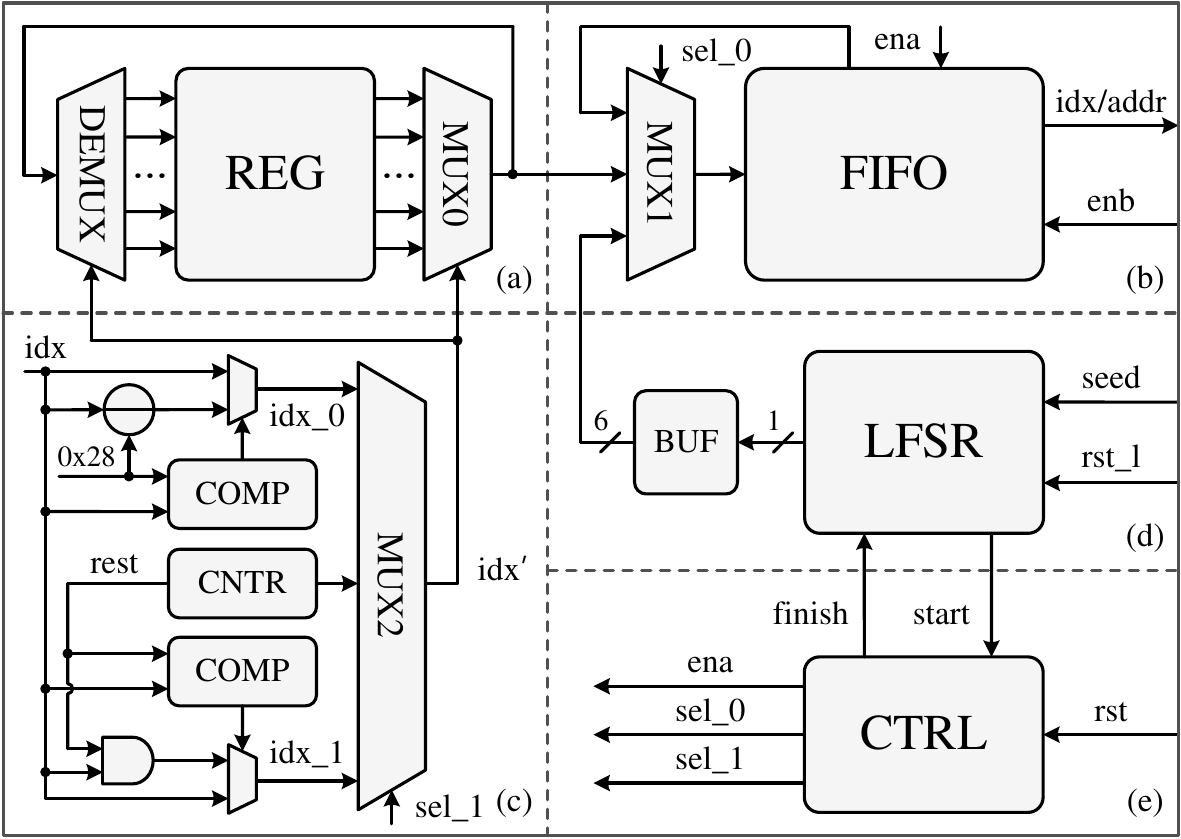}
\caption{The proposed architecture of RPG.} 
\label{RPG}
\end{figure}

Based on the above analysis, the proposed architecture of RPG is shown in Fig.~\ref{RPG}, divided into five parts. Part (a) contains one 64-to-1 MUX named MUX0, one 1-to-64 DEMUX, and one register group named REG with data width of 6 and depth of 64. REG stores the designed address permutation and its updated version. Under the control of $idx'$, MUX0 outputs an element from REG continuously and DEMUX outputs an element from MUX0 to update the data in REG. Part (b) contains one 3-to-1 MUX and a FIFO with data width of 6 and depth of 64, which is reused to cache the random indexes from the LFSR module and the generated random address permutation to reduce the area consumption. Three control signals are used, \textit{i.e.}, $sel\_0$, $ena$, and $enb$. With the control signals of $ena$ and $enb$, FIFO is worked in the input mode and the cyclic mode, respectively. When FIFO is in the input mode, the data enter from REG or LFSR; and when FIFO is in the cyclic mode, the output is pushed back into the input. Part (c) contains one 3-to-1 MUX named MUX2, two 2-to-1 MUXs, two comparators, one counter, one subtractor, and one AND gate. MUX2 has three inputs, $idx\_0$, $idx\_1$, and $rest$, controlled by $sel\_1$. The input random index $idx$ output from FIFO is adjusted to match the two stages of shuffling, formulated as:
\begin{equation}
\label{10}
idx\_0=
\left\{
\begin{aligned}
&idx \ (idx\leq28), \\
&idx-28 \ (idx>28),
\end{aligned}
\right.
\end{equation}
\begin{equation}
\label{11}
idx\_1=
\left\{
\begin{aligned}
&idx \ (idx\leq rest), \\
&idx \ \& \ rest \ (idx>rest),
\end{aligned}
\right.
\end{equation}
where the value 28 is in the hexadecimal format. The variable \textit{rest} denotes the number of the remaining elements to be selected minus 1, which is computed by a decrement counter. The output signal $idx'$ is selected from $idx\_0$, $idx\_1$, and $rest$,  served as the control signal of MUX0 and DEMUX in (a). Part (d) contains a linear feedback shift register (LFSR) with depth of 32 and a buffer. The buffer outputs every six cycles to convert six 1-bit numbers into a 6-bit number (index). The random seeds are generated by an external true random number generator (TRNG). When $rst\_l$ is equal to 1, the LFSR will be updated by the input random seed. Part (e) contains a controller made up with several logic circuits. It produce the control signals $ena$, $sel\_0$, $sel\_1$, and $finish$ based on the input signals for the other three modules.

\begin{algorithm}[!t]
\caption{The Processing Schedule of RPG}\label{alg1}
\footnotesize
\begin{algorithmic}
\STATE
\STATE 1:~\textbf{REG}$\leftarrow \{0b,0c,...,32,33,3f,3e,...,3a,39,$
\STATE ~~~~~~~~~~~~~~$0a,09,...,01,00,38,37,36,35,34\}$
\STATE 2:~\textbf{FIFO$_{ena}$}$\leftarrow 1$
\STATE 3:~\textbf{for} $i=0$ to $63$ \textbf{do}
\STATE 4:~~~~~\textbf{for} $j=0$ to $5$ \textbf{do}
\STATE 5:~~~~~~~~~\textbf{BUF}$[j]$ $\leftarrow$ \textbf{LFSR}
\STATE 6:~~~~~\textbf{end for}
\STATE 7:~~~~~\textbf{FIFO$_{in}$} $\leftarrow$ \textbf{BUF}
\STATE 8:~\textbf{end for}
\STATE 9:~\textbf{for} $k=0$ to $11$ \textbf{do}
\STATE \hspace{-0.17cm}10:~~~~~$rest\leftarrow 63-k$
\STATE \hspace{-0.17cm}11:~~~~~$idx\_0\leftarrow$ \textbf{FIFO$_{out}$}$>0x28$ ? \textbf{FIFO$_{out}$} : \textbf{FIFO$_{out}$} $- \ 0x28$
\STATE \hspace{-0.17cm}12:~~~~~\textbf{FIFO$_{in}$}$\leftarrow$ \textbf{REG}$[idx\_0]$
\STATE \hspace{-0.17cm}13:~~~~~\textbf{REG}$[idx\_0]$$\leftarrow$ \textbf{REG}$[rest]$
\STATE \hspace{-0.17cm}14:~\textbf{end for}
\STATE \hspace{-0.17cm}15:~\textbf{for} $k=12$ to $63$ \textbf{do}
\STATE \hspace{-0.17cm}16:~~~~~$rest\leftarrow 63-k$
\STATE \hspace{-0.17cm}17:~~~~~$idx\_1\leftarrow$ \textbf{FIFO$_{out}$}$>rest$ ? \textbf{FIFO$_{out}$} : \textbf{FIFO$_{out}$} $\& \ rest$
\STATE \hspace{-0.17cm}18:~~~~~\textbf{FIFO$_{in}$}$\leftarrow$ \textbf{REG}$[idx\_1]$
\STATE \hspace{-0.17cm}19:~~~~~\textbf{REG}$[idx\_1]$$\leftarrow$ \textbf{REG}$[rest]$
\STATE \hspace{-0.17cm}20:~\textbf{end for}
\STATE \hspace{-0.17cm}21:~\textbf{for} $i=0$ to $5$ \textbf{do}
\STATE \hspace{-0.17cm}22:~~~~~\textbf{FIFO$_{in}$}$\leftarrow$ \textbf{FIFO$_{out}$}
\STATE \hspace{-0.17cm}23:~\textbf{end for}
\STATE \hspace{-0.17cm}24:~\textbf{FIFO$_{ena}$}$\leftarrow 0$
\end{algorithmic}
\end{algorithm}

To make it more clear, we give an algorithm to illustrate the processing schedule of RPG as shown in Algorithm~\ref{alg1}. The generation of random permutations can be divided into four steps: (a) initialization, (b) shuffling\_12, (c) shuffling\_52, and (d) cyclic shift\_6. In the initialization step, LFSR outputs $6\times 64=384$ bits to fill up FIFO, and REG is set with the designed address permutation. Note that the 64 registers in REG use $00$ to $3f$ as their serial numbers. In the shuffling\_12 step, the 12 random addresses are selected from the 41 ($0x00\sim 0x28$) lower-side locations of REG based on the shift-out elements from FIFO$_{out}$ and the equation~\eqref{10}, and cached into FIFO$_{in}$. Meanwhile, those selected locations of REG are filled up with its elements from the specific locations (corresponding to the dynamic maximum serial number $rest$). After 12 rounds, this step is finished and the shuffling\_52 step starts. In this step, the rest 52 addresses in REG are messed up based on the remaining random indexes in FIFO and the equation~\eqref{11}. They are serially chosen and cached into FIFO from the lower-side locations of REG. Assume that when $rest$ is $0x02$, the random index is $0x05$, larger than $0x02$. When it is computed based on the equation~\eqref{11}, we get the adjusted index equal to $0x00$. We then use the new index to choose the element in the zeroth register of REG and push it into FIFO from the lower-side. At the same time, the element in the zeroth register is updated by the element in the second register (corresponding to the dynamic maximum serial number $rest$) of REG. When the shuffling\_52 step is completed, we start the fourth step, \textit{i.e.}, the cyclic shift\_6 step. FIFO is shifted six times in the cyclic mode and we get the target random address permutation. 

The protection will not increase the consumption of total time of Kyber since RPG is conducted in parallel with Kyber, costs much fewer cycles, and has a shorter critical path.

\begin{figure}[!t]
\centering
\includegraphics[width=3.33in]{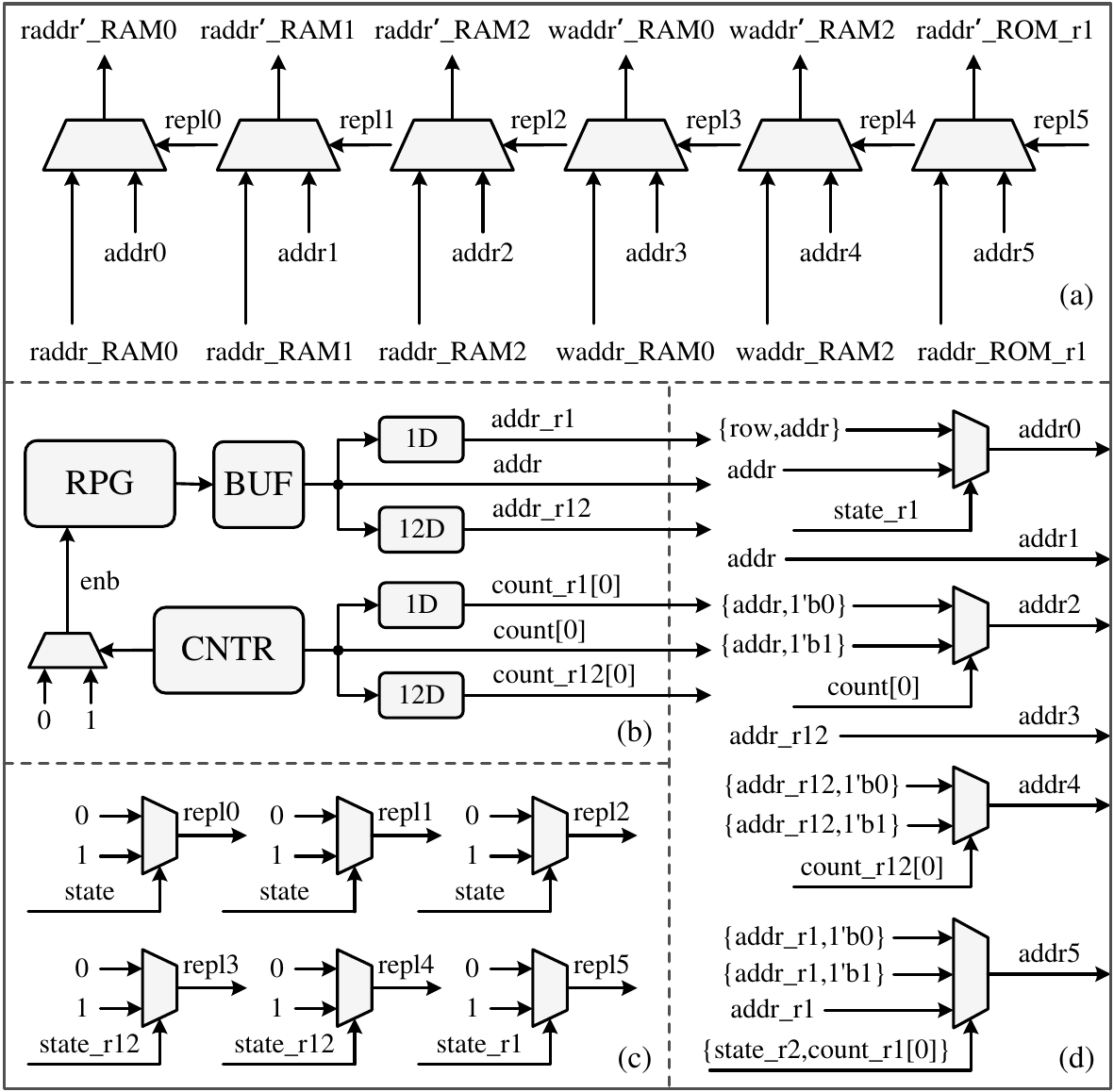}
\caption{The proposed architecture of ADDR.}
\label{addr}
\end{figure}

\begin{table*}[b]
\begin{threeparttable}
\caption{Comparison of the FPGA Area and Time Performance}\label{tab:fpga}
\centering
\renewcommand{\arraystretch}{1.6} 
\setlength{\tabcolsep}{6.3pt} 
\begin{tabular}{|c|c|c|c|c|c|c|c|c|c|c|c|}
\hline
Implementation & Parameter & \makecell{Platform} & LUTs & FFs & Slices & DSPs & BRAMs & ENS $^1$ & \makecell{Cycles \\ ($\times 10^3$)} & \makecell{Frequency \\ (MHz)} & \makecell{ATP $^2$ \\ (ENS$\times ms$)}\\
\hline
Xing \textit{et al}.~\cite{xing2021compact}$^3$ & Kyber512 & Artix-7 & 7353 & 4633 & 2173 & 2 & 3 & 2973 & 6.7 & 206 & 96.7\\
\hline
Kamucheka \textit{et al}.~\cite{kamucheka2022masked} & Kyber512 & Virtex-7 & 143112 & - & 81746 & 60 & 294 & 146546 & 126.6 & 100 & 186$\times 10^3$\\
\hline
Jati \textit{et al}.~\cite{jati2024configurable} & Kyber512 & Artix-7 & 7151 & 3730 & 2260 $^4$ & 2 & 5.5 & 3560 & 57.2 & 258 & 789.3\\
\hline
\textbf{This work} & Kyber512 & Artix-7 & \textbf{8143} & \textbf{5151} & \textbf{2433} & \textbf{2} & \textbf{3} & \textbf{3233} & \textbf{6.7} & \textbf{206} & \textbf{105.2}\\
\hline
Xing \textit{et al}.~\cite{xing2021compact}$^3$ & Kyber768 & Artix-7 & 7353 & 4633 & 2173 & 2 & 3 & 2973 & 10.0 & 206 & 144.3\\
\hline
Moraitis \textit{et al}.~\cite{moraitis2023securing} & Kyber768 & Artix-7 & 14341 & 9190 & 4734 $^4$ & $\geq$2 & 6 & $\geq$6134 & 10.0 & $\leq$206 & $\geq$297.8\\
\hline
\textbf{This work} & Kyber768 & Artix-7 & \textbf{8143} & \textbf{5151} & \textbf{2433} & \textbf{2} & \textbf{3} & \textbf{3233} & \textbf{10.0} & \textbf{206} & \textbf{156.9}\\
\hline
\end{tabular} 
\begin{tablenotes}
\item 1. ENS (equivalent number of slices) = \#Slices + 100 $\times$ \#DSPs + 200 $\times$ \#BRAMs ~\cite{li2022reconfigurable}.
\item 2. ATP (area-time product) = ENS $\times$ Time ($ms$) = ENS $\times$ Cycles $\times$ 1/Frequency $\times$ $10^3$.
\item 3. The unprotected work was originally evaluated using Artix-7 XC7A12T. We reevaluate it using Artix-7 XC7A100T for a fair comparison.
\item 4. The number of Slices is approximately computed by using 0.25 $\times$ \#LUTs + 0.125 $\times$ \#FFs~\cite{li2022reconfigurable}.
\end{tablenotes}
\end{threeparttable}
\end{table*}

\begin{figure}[!t]
\centering
\includegraphics[width=3.5in]{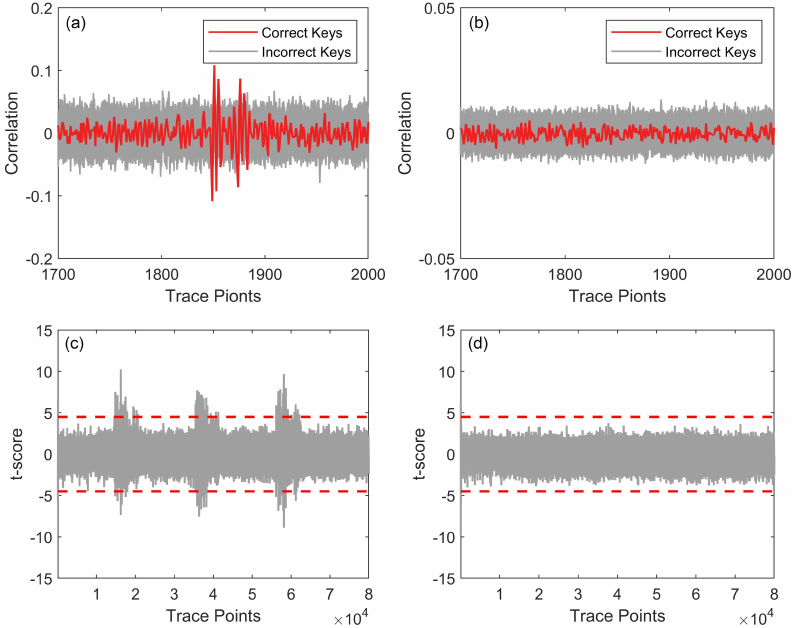}
\caption{The CPA results of the unprotected scheme with $4\times 10^3$ EM traces (a) and the protected scheme with $1\times 10^5$ EM traces (b). The TVLA results of the unprotected scheme with $1\times 10^4$ EM traces (c) and the protected scheme with $1\times 10^7$ EM traces (d). The red dashed lines represent $\pm4.5$.}
\label{evaluation}
\end{figure}

\subsection{Address Controller}
As shown in Fig.~\ref{addr}, the proposed architecture of the address controller (ADDR) is divided into four parts. In part (a), four new read addresses (raddr'\_x) and two new write addresses (waddr'\_x) are used to replace the read and write addresses of the RAMs and ROM. Six 2-to-1 MUXs are used to determine the data sources of these new addresses. When there is no need for shuffling, the original addresses are directly connected. When shuffling is required, $addr0\sim addr5$ are selected. The generation of the control signals $repl0\sim repl5$ for these MUXs is shown in part (c). In part (b), four shift registers are used to output the delayed random permutations and control signals. The external enable signal $enb$ of FIFO in RPG is set to 1 every other clock cycle (PWM and subtraction). The generation of $addr0\sim addr5$ is shown in part (d). The permutations $40\sim7f$ and $00\sim7f$ required by $addr2$, $addr4$, and $addr5$ can be obtained by concatenating one 0 or 1 after $addr$, $addr\_r12$, and $addr\_r1$, respectively. The four permutations $00\sim3f$, $40\sim7f$, $80\sim bf$, and $c0\sim ff$ required by $addr0$ are obtained by concatenating $row$ (equal to $0\sim 3$) and $addr$ into $\{row,addr\}$. 

\section{Results and Comparisons}
\subsection{Evaluation on SCA Resistance}
To better assess the level of improvement in side-channel security after applying the protective measures, we conduct correlation power analysis (CPA) on PWM and test vector leakage assessment (TVLA) on Kyber's decryption procedure. In the experiments, Pearson correlation coefficient and Welch's t-test are used as metrics in CPA and TVLA, respectively. We deploy our experiments for Kyber768 on a target board containing an Artix-7 XC7A100T FPGA. 

We target the registers at the output ports of multipliers and utilize the Hamming distance model to simulate the energy consumption resulting from the dynamic flipping of these registers, similar to the methodology employed in~\cite{zhao2023side}. Finally, the correlation between the assumed energy values and the collected energy values are calculated to analyze the correct keys. As shown in Fig.~\ref{evaluation}, all experiments have achieved the expected results, \textit{i.e.}, the proposed shuffling countermeasure with huge permutation space can significantly improve the side-channel security of Kyber's hardware implementation. It should be pointed out that we have omitted the CPA results of modular reduction and subtraction here for simplicity as they adopt the same defense countermeasure and have the similar conclusions.

\subsection{Comparisons on Area and Performance}
Table~\ref{tab:fpga} shows the experiment results and comparison of Kyber's hardware implementations with different hiding schemes, including the area and time performance. Compared to the unprotected Kyber implementation~\cite{xing2021compact}, the proposed protected design only consumes an extra resource of 260 Slices. The numbers of DSPs and BRAMs both are the same. Compared with the unprotected design, the ATP (also area) of the proposed version is increased by only 8.7\%.

We also make comparisons with previous protected Kyber hardware implementations. It can be seen that the hiding design proposed in~\cite{kamucheka2022masked} has tremendous area and time consumption, more than three orders of magnitude larger than ours in ATP. The design in~\cite{jati2024configurable} has a relatively small area but consumes huge cycles, so its ATP is also much worse than ours. In contrast, the design in~\cite{moraitis2023securing} has the same clock cycles, but their area is almost twice that of ours.

\section{Conclusion}
In this brief, we have devised an optimized shuffling architecture against SCAs for Kyber's hardware implementation. The experimental results show that the proposed design can effectively improve the side-channel security, and reports only 8.7\% degradation on the hardware efficiency compared with the original unprotected version, much better than existing hiding schemes.

\section{Discussion}
For lattice-based KEMs and signature algorithms, NTT-based polynomial multiplications serve as their most basic operation, with similar security foundations, making the proposed method applicable to other lattice-based schemes. In addition, guarding against FIAs is equally important as guarding against SCAs. However, this brief has so far only protected the Kyber algorithm and has not yet integrated error detection schemes. Future work will apply the proposed method to other promising algorithms like Dilithium and Raccoon, and consider combining it with error detection schemes.

\end{document}